\documentclass{elsart}

\usepackage{graphicx}
\usepackage{amsmath}
\usepackage{amssymb}

\begin{document}

\begin{frontmatter}

\title{Lifetime of $^{26}$S and a limit for its $2p$ decay energy}

\author[dub]{A.S.~Fomichev,}
\author[gsi,val]{I.G.~Mukha,}
\author[dub]{S.V.~Stepantsov,}
\author[dub,gsi,kur]{L.V.~Grigorenko,}
\author[gsi,unifrankfurt]{E.V.~Litvinova,}
\author[dub,opa]{V.~Chudoba,}
\author[bog]{I.A.~Egorova,}
\author[dub]{M.S.~Golovkov,}
\author[dub]{A.V.~Gorshkov,}
\author[dub]{V.A.~Gorshkov,}
\author[dub,pan]{G.~Kaminski,}
\author[dub]{S.A.~Krupko,}
\author[dub,msu]{Yu.L.~Parfenova,}
\author[dub]{S.I.~Sidorchuk,}
\author[dub]{R.S.~Slepnev,}
\author[dub]{G.M.~Ter-Akopian,}
\author[dub,pan]{R.~Wolski,}
\author[cha]{M.V.~Zhukov}

\address[dub]{Flerov Laboratory of Nuclear Reactions, JINR, RU-141980 Dubna,
Russia}

\address[gsi]{GSI Helmholtzzentrum f\"{u}r Schwerionenforschung,
Planckstra{\ss}e 1, D-64291 Darmstadt, Germany}

\address[val]{CSIC-IFIC, University of Valencia, E-46071 Valencia, Spain}

\address[kur]{Russian Research Center ``The Kurchatov Institute'', Kurchatov
sq.\ 1, RU-123182 Moscow, Russia}

\address[unifrankfurt]{Institut f\"{u}r Theoretische Physik,
Goethe-Universit\"{a}t, 60438 Frankfurt am Main, Germany}

\address[opa]{Institute of Physics, Silesian University in Opava, Bezru\v{c}ovo
n\'{a}m.\ 13, 74601 Czech Republic}

\address[bog]{Bogolubov Laboratory of Theoretical Physics, JINR, Dubna,
RU-141980 Russia}

\address[pan]{Institute of Nuclear Physics PAN, Radzikowskiego 152, PL-31342
Krak\'{o}w, Poland}

\address[msu]{Skobel'tsyn Institute of Nuclear Physics, Moscow State University,
119991 Moscow, Russia}

\address[cha]{Fundamental Physics, Chalmers University of Technology, S-41296
G\"{o}teborg, Sweden}

\begin{abstract}
Unknown isotope $^{26}$S, expected to decay by two-proton ($2p$) emission, was 
studied theoretically and was searched experimentally. The structure of this 
nucleus was examined within the relativistic mean field (RMF) approach. A method 
for taking into account the many-body structure in the three-body decay 
calculations was developed. The results of the RMF calculations were used as an 
input for the three-cluster decay model worked out to study a possible $2p$ 
decay branch of this nucleus. The experimental search for $^{26}$S was performed 
in fragmentation reactions of a 50.3 A MeV $^{32}$S beam. No events of $^{26}$S 
or $^{25}$P (a presumably proton-unstable subsystem of $^{26}$S) were observed. 
Based on the obtained production systematics an upper half-life limit of 
$T_{1/2}<79$ ns was established from the time-of-flight through the fragment 
separator. Together with the theoretical lifetime estimates for two-proton decay 
this gives a decay energy limit of $Q_{2p}>640$ keV for $^{26}$S. Analogous 
limits for $^{25}$P are found as $T_{1/2}<38$ ns and $Q_{p}>110$ keV. In the 
case that the one-proton emission is the main branch of the $^{26}$S  decay a 
limit $Q_{2p}>230$ keV would follow for this nucleus. It is likely that $^{26}$S 
resides in the picosecond lifetime range and the further search for this isotope 
is prospective for the decay-in-flight technique.
\end{abstract}


\end{frontmatter}


\section{Introduction}

\begin{figure}
\centerline{
\includegraphics[width=0.8 \textwidth]{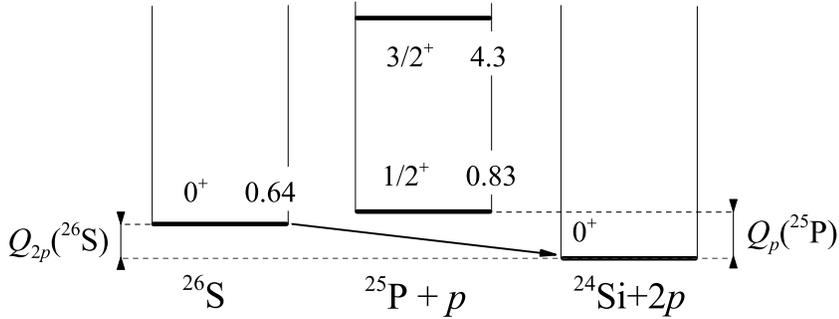}  
}
\caption{Energy levels and decay scheme for a possible true $2p$ decay of
$^{26}$S. The threshold energies are taken from systematics \cite{audi}.}
\label{fig:levels}
\end{figure}

In the recent years there were significant advances in the studies of proton 
dripline nuclei. The phenomenon of true two-proton ($2p$) radioactivity (true 
three-body decay) proposed in Ref.\ \cite{gol60} is an important object of these 
studies. The $2p$ radioactivity is an exclusive quantum-mechanical phenomenon 
occurring under specific decay energy conditions ($Q_{2p}>0$, $Q_p<0$) which 
make the sequential proton emission impossible, see Fig.\ \ref{fig:levels}. Both 
protons in this case should be emitted simultaneously. The necessary energy 
conditions are widespread near the proton dripline where they are connected to 
the pairing correlations (the effect takes place in isotopes with even numbers 
of protons). A consistent quantum mechanical theory of the $2p$ radioactivity 
based on a three-body cluster model was for the first time proposed in 
\cite{gri00b}, and further developed in \cite[and Refs.\ 
therein]{gri01a,gri03c,gri07,gri09,gri10}. Being the latest discovered mode of 
radioactive decay (discovered in 2002 in $^{45}$Fe \cite{pfu02,gio02}) the $2p$ 
radioactivity was soon found in several other nuclei $^{54}$Zn \cite{bla05}, 
$^{48}$Ni \cite{dos05}, $^{19}$Mg \cite{muk07}. The two-proton ground-state 
decays were recently investigated in detail for relatively short-lived nuclei 
$^{6}$Be \cite{gri09} and $^{16}$Ne \cite{muk07}. The modern trend of this 
research lies in the studies of the proton-proton ($p$-$p$)  correlations 
\cite{muk07,mie07,gri09}. These correlations are expected to be closely 
connected to the structure of the $2p$ emitters \cite{gri03c,gri09}. Searches 
for other $2p$-radioactivity cases were proposed in $^{30}$Ar, $^{34}$Ca 
\cite{gri03}, $^{62}$Se, $^{58,59}$Ge, $^{66,67}$Kr \cite{gri03c}. Several 
experiments dedicated to this phenomenon in $^{6}$Be, $^{48}$Ni, and $^{30}$Ar 
are now under the way.

\begin{table}[bt]
\caption{Energies and charge radii of $^{24}$Si, $^{25}$P, and $^{26}$S nuclei 
relevant to the $2p$-decay problem. The energies are given in MeV and the radii 
are given in fm. The phenomenological pairing gap is defined as 
$\Delta_{\text{pg}}=2Q_p(^{25}$P$)-Q_{2p}(^{26}$S). The corrected $2p$-decay 
energy $Q^{\text{corr}}_{2p}(^{26}$S$)=\Delta_{\text{cde}}-S_{2n}(^{26}$Ne) is 
based on the calculated Coulomb displacement energy $\Delta_{\text{cde}}$ and on 
the experimental two-neutron separation energy in the mirror system $^{26}$Ne. 
Maximal and minimal predictions are listed for papers providing several sets of 
results. The two three-body model (3b) results with different 
$Q^{\text{corr}}_{2p}$ values are obtained with different charge distributions 
in $^{24}$Si. The results of systematics studies are collected in the upper part 
of the Table and are separated from calculations by a horizontal line.}
\vspace{2mm}
\begin{center}
\tabcolsep=0.5em \renewcommand{\arraystretch}{1.0}
\begin{tabular}[c]{ccccccc}
\hline \hline
Ref.  & $Q_{2p}(^{26}$S) & $Q_{p}(^{25}$P) & $\Delta_{\text{pg}}$  &
$r_{\text{ch}}(^{24}$Si) &
$r_{\text{ch}}(^{26}$S) & $Q^{\text{corr}}_{2p}$  \\
\hline
\cite{audi}   & 0.637 & 0.826 & 1.01 &  &  &   \\
\cite{col98}  & 2.116 & 1.516 & 0.92 & & & \\
              & 1.954 & 1.384 & 0.81 & & & \\
\hline
\cite{pat97}  & 1.784 & &   & 3.21 & 3.37 & 1.61  \\
              & 0.763 & &   & 3.25 & 3.44 & 0.43   \\
\cite{lal98}  & 0.500 &       &       & 3.24 & 3.27 & 1.28 \\
\cite{che98}  & 0.585 & 0.325 & 0.07 & 3.05  &       & \\
              & 0.013 & 0.019 & 0.02 & 3.05  &       & \\
\cite{wan07}  &       &       &       & 3.24 & 3.49 & \\
 RMF & 0.47 & 1.40    & 2.32  & 3.71  & 3.76& 1.78 \\
 3b        &    &  0.83  &  &  3.71 &       &  1.1 \\
           &    &  0.83  &  &  3.05 &       &  0.8 \\
\hline \hline
\end{tabular}
\end{center}
\label{tab:s26-info}
\end{table}

The system $^{26}$S remained somehow ``in shadow'' while the interest of
researchers were focused on other nuclei. Very little can be found about
$^{26}$S in literature. The Brookhaven database \cite{nndc} assigns a half-life
of $T_{1/2} \approx 10$ ms and a possible $2p$ decay mode to this nucleus.
Unfortunately, the source of this information is not clear. The standard
compilation \cite{audi} indicates that $^{26}$S is likely to be nuclear unstable
[$Q_{2p}(^{26}$S$)>0$] and is likely to be a true two-proton emitter
[$Q_p(^{25}$P$)>Q_{2p}(^{26}$S$)$], see Table \ref{tab:s26-info}.

The $^{26}$S nucleus has never been studied theoretically in details. It has 
been briefly mentioned in papers \cite{bro96,col98} (phenomenology) as a 
possible $2p$ emitter. RMF calculations were performed in 
\cite{pat97,lal98,che98,wan07}. The results of these studies for $^{26}$S are 
quite controversial (information relevant to $^{26}$S and its subsystems is 
summarized in Table \ref{tab:s26-info}). The predicted $2p$-decay energies are 
scattered in a broad range between 0 and 2 MeV, and the radial characteristics 
vary considerably as well. The overall situation is such that the existing 
theoretical information is not very helpful for planning experimental search for 
$^{26}$S. It is not even clear if $^{26}$S is a true two-proton emitter 
[$Q_p(^{25}$P$)>Q_{2p}(^{26}$S$)$] or sequential one-proton emission is possible 
for this nucleus [$S_p(^{26}$S$)<0$]. The lifetime vs decay energy systematics 
are drastically different in these cases. There are theoretical works supporting 
both possibilities and absolutely no experimental data.

\begin{figure}
\centerline{
\includegraphics[width=0.8 \textwidth]{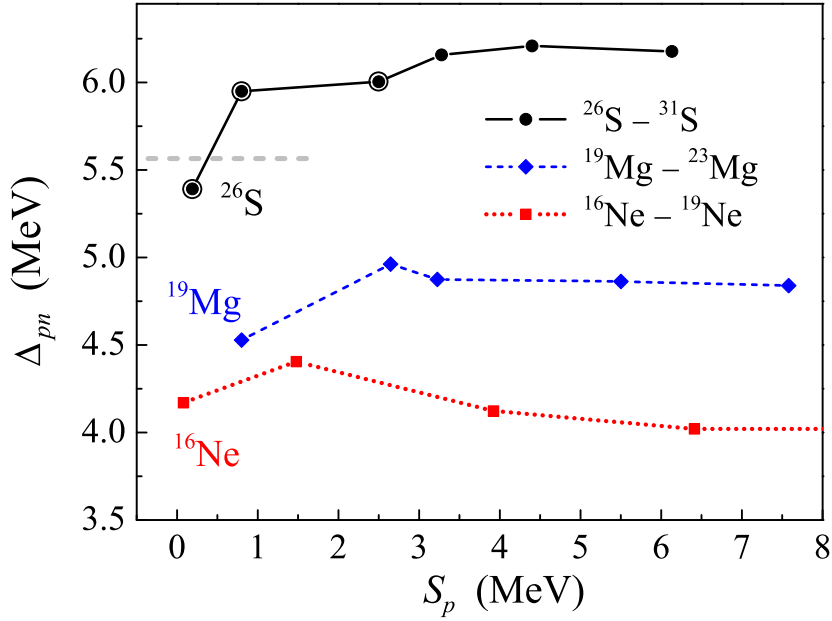}
}
\caption{Systematics $\Delta_{pn}$ vs $S_p$. For the neon and magnesium chains 
the values are experimental; for the sulphur chain they are partly taken from 
systematics Ref.\ \cite{audi} (dots with circles around). Below the gray dashed 
horizontal line the $S_p$ value for $^{26}$S is positive (it is a true 
two-proton emitter) and above it is negative (the sequential one-proton decay 
becomes possible).}
\label{fig:deltapn}
\end{figure}

We would like to provide some support for the true $2p$ character of the 
$^{26}$S decay using the recent experimental data on $^{19}$Mg \cite{muk07}. A 
negative value of $S_p=-0.42$ MeV was inferred for $^{26}$S in Ref.\ 
\cite{bro96} basing on the $\Delta_{pn}$ vs $S_p$ systematics. Here 
$\Delta_{pn}(Z)=S_n(^A_ZN)-S_p(^A_NZ)$ is the difference of the nucleon 
separation energies in isobaric mirror partners. It was assumed in Ref.\ 
\cite{bro96} that the $\Delta_{pn}$ behaviour should be smooth and thus a value 
of $\Delta_{pn} \sim 6$ MeV was estimated for $^{26}$S (in contrast with 
$\Delta_{pn} = 5.39$ predicted by \cite{audi}). This systematics is shown in 
Fig.\ \ref{fig:deltapn} but together with the same ones for the neon and 
magnesium chains. One can easily see that for the \emph{known} true $2p$ 
emitters $^{16}$Ne and $^{19}$Mg, also belonging to the $s$-$d$ shell, there is 
a break in the smooth trend of $\Delta_{pn}$. Such a break could be connected, 
for example, to structural peculiarities of the three-cluster systems in 
proximity to the proton dripline \cite{gri02}. The $\Delta_{pn}$ value is 
shifted downwards by 0.35-0.5 MeV for the known neighbouring systems when we 
pass the proton dripline. This observation is in a good agreement with the 
prediction of Ref.\ \cite{audi} for $^{26}$S supporting the idea about the true 
$2p$ character of the $^{26}$S decay.

The discussed situation motivated us to make first a dedicated
theoretical study of $^{26}$S focusing on the possibility of true two-proton
emission and then to perform an experimental search for $^{26}$S if the
theoretical results look promising.

\section{The many-body structure aspect of the $2p$ decay}


Our approach to the $2p$ radioactivity problem \cite{gri00b} is based on the
approximate boundary conditions for the Coulomb three-body problem formulated in
Ref.\ \cite{gri01a}. The three-body wave function (WF)
$\Psi_3^{(+)}(\rho,\Omega_5)$ with pure outgoing boundary conditions and complex
energy $E=Q_{2p}-i\Gamma/2$ is constructed for an arbitrary small value of
$\Gamma$ using the hyperspherical harmonic method. In this method the WF depends
on the collective variable hyperradius $\rho$ and on the 5-dimensional
hyperangle $\Omega_5$. After that the actual $\Gamma$ is found using the
``natural'' definition
\begin{equation}
\Gamma= \frac{j}{N}=\frac{\text{Im} \int \! d \Omega_5 \left.
\Psi^{(+)\dagger}_3 \rho^{5/2} (d/d\rho) \rho^{5/2} \Psi^{(+)}_3 \right|
_{\rho_{\max}} } {||\Psi_3^{(+)}||^2} \, .
\label{eq:width3}
\end{equation}
The differentials of the flux $j$ defined on the hypersphere with large
hyperradius $\rho_{\max}\sim 10^3$ fm and the classical trajectory extrapolation
to $\rho_{\text{ext}}\sim 10^5$ fm are used to define the momentum
distributions (see Refs.\ \cite{gri03c,gri07,gri10} for technical details).

The three-cluster approach to the $2p$ decays is well justified  for the 
closed-shell systems or systems with a closed-shell core. However, most of the 
two-proton emitters, including $^{26}$S, do not belong to this class, and 
important effects of the many-body structure can be expected for them. The major 
components of the three-body cluster WFs with $J^{\pi}=0^+$ can be written in a 
schematic spectroscopic notation as
\begin{equation}
\Psi_3^{(+)}   = \sum \nolimits _i X_i [l_i^2]_0  \quad ; \qquad
\sum \nolimits _i |X_i|^2 \lesssim 1 \,.
\label{eq:wf3}
\end{equation}
The schematic notation $[l_{i}^n] _J$ denotes the Slater determinant of $n$
nucleons occupying $l_{i}$ orbital projected on the total spin $J$ and
normalized: the three-body WF can be considered as normalized in the internal
region
\begin{equation}
||\Psi_3^{(+)}||^2 =\int d \Omega_5 \int_0^{\rho_{\max}} d \rho \, \rho^5
|\Psi_3^{(+)}|^2 \equiv 1
\label{eq:label}
\end{equation}
without loss of generality. In papers \cite{gri03c,gri07} the results of the
$2p$ width calculations were provided as functions of weights of the main
cluster configurations $w(l^2_i)=X_i^2$. Therefore we can write
\begin{equation}
\Gamma(X^2_i)= j(X^2_i) \,.
\label{eq:gamma-j}
\end{equation}
The values which can be put in correspondence with the components of the
three-body WF (\ref{eq:wf3}), to take into account the many-body structure, are
overlaps of the many-body WFs of the precursor-daughter pair multiplied by a
combinatorial term
\begin{equation}
\Psi_{A,A-2} =  \left( \frac{A!}{2!(A-2)!} \right) ^{1/2} \langle \Psi_{A} |
\Psi_{A-2} \rangle = \sum _i \tilde{X}_i [l_i^2]_0 \,.
\label{eq:psi-over}
\end{equation}
The overlaps of Eq.\ (\ref{eq:psi-over}) are in general case not normalized to
unity:
\begin{equation}
N_{2p} = \sum \nolimits _i |\tilde{X}_i|^2 \neq 1 \,.
\label{eq:norm}
\end{equation}
If we normalize them to unity, $\Psi_{A,A-2} \rightarrow \Psi_{A,A-2}
N_{2p}^{-1/2}$, then the initial many-body WF becomes not normalized to unity
$||\Psi_{A}|| =N_{2p}^{-1}$. According to Eq.\ (\ref{eq:width3}) the width
should be then renormalized as
\begin{equation}
\tilde{\Gamma} = N_{2p} \, \Gamma( \tilde{X}^2_i/N_{2p} ) \,.
\label{eq:width-mb}
\end{equation}
The  renormalized width contains a product with structural factor ($N_{2p}$)
which can be easily interpreted as a kind of a spectroscopic factor in analogy
with the two-body decays. However, contrary to the two-body case, the
dependence of the width on the structure includes the dependence on the
amplitudes $X_i$. They should be adjusted as $X_i \rightarrow \tilde
X_i/\sqrt{N_{2p}}$. That makes the above renormalization a complicated nonlinear
procedure. We will see below that the many-body effects lead to a significant 
renormalization of the width beyond the Hartree approximation.

\begin{figure}
\centerline{
\includegraphics[width=0.9 \textwidth]{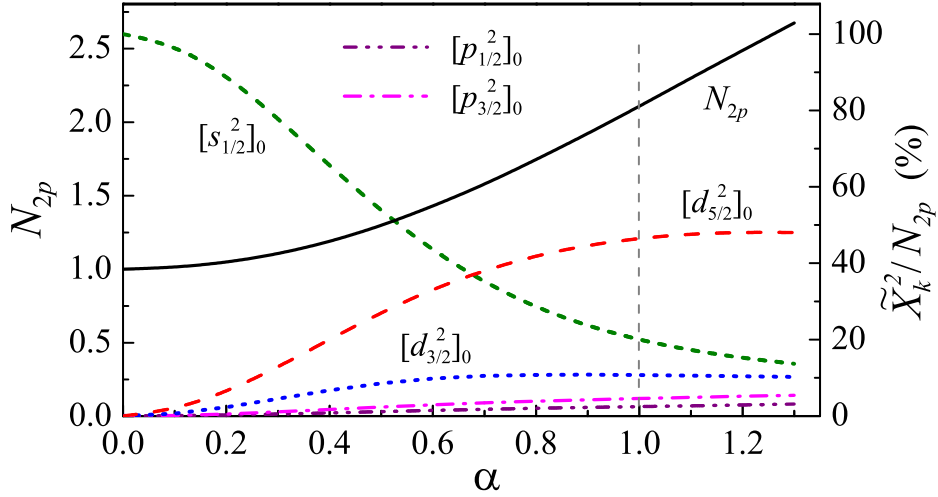}  
}
\caption{(Colour online) Weights of the major components of the overlaps Eq.\
(\ref{eq:psi-over}) (left axis) and the total overlap normalization as a
function of the parameter $\alpha$ (right axis). This parameter scales the
pairing gap $\Delta \rightarrow \alpha \Delta$.}
\label{fig:uv-evol}
\end{figure}

To calculate the amplitudes $\tilde{X}_i$ in a many-body approach we can use a
formalism analogous to the same intrinsic structure model as for direct
two-nucleon transfer reactions [for instance, pick-up processes ($p,t$),
($n$,$^3$He) etc]. In the present work we consider the following situation: (i)
both the parent nucleus $Z$ and the daughter nucleus $Z-2$ are essentially
many-body systems, (ii) both nuclei are spherical in their ground states, (iii)
the emitted proton pair transfers zero total angular momentum, (iv) both
nuclei remain in their ground states.

The structure part of the two-nucleon transfer cross section is determined by
the spectroscopic amplitude $B$ which is defined in the second quantization
formalism as \cite{yos62,bro67}:
\begin{equation}
B_{(K_i K_f,k k^{\prime})}^{J} = \sum \limits_{MM_i} C_{J_iM_iJM}^{J_fM_f}
\langle \Psi_{K_f}| \mathcal{O}_{(k k^{\prime})}^{JM} |\Psi_{K_i}\rangle,
\label{eq:sa}
\end{equation}
where $K_i = \{(K_i),M_i\}$ and $K_f = \{(K_f),M_f\}$ denote quantum number sets
of the initial and final nuclei with $\bigl( K_i; K_f \bigr)$ and without
$\bigl( (K_i); (K_f) \bigr)$ angular momentum projections $M_i$, $M_f$. Index
$k$ denotes the full set of the single-particle quantum numbers $k = 
\{(k),m_k\}$ in a spherical nucleus. $\mathcal{O}$ is the two-nucleon transfer 
operator:
\begin{equation}
\mathcal{O}_{(k k^{\prime})}^{JM} =  \frac{1}{1 + \delta_{(kk^{\prime})}}
\sum \limits_{m_km_{k^{\prime}}} C_{j_km_kj_{k^{\prime}}m_{k^{\prime}}}^{JM}
a_k a_{k^{\prime}} \,,
\label{eq:tnto}
\end{equation}
where $a_k$ is the nucleon annihilation operator in the mean field basis.

\begin{figure}
\centerline{
\includegraphics[width=0.9 \textwidth]{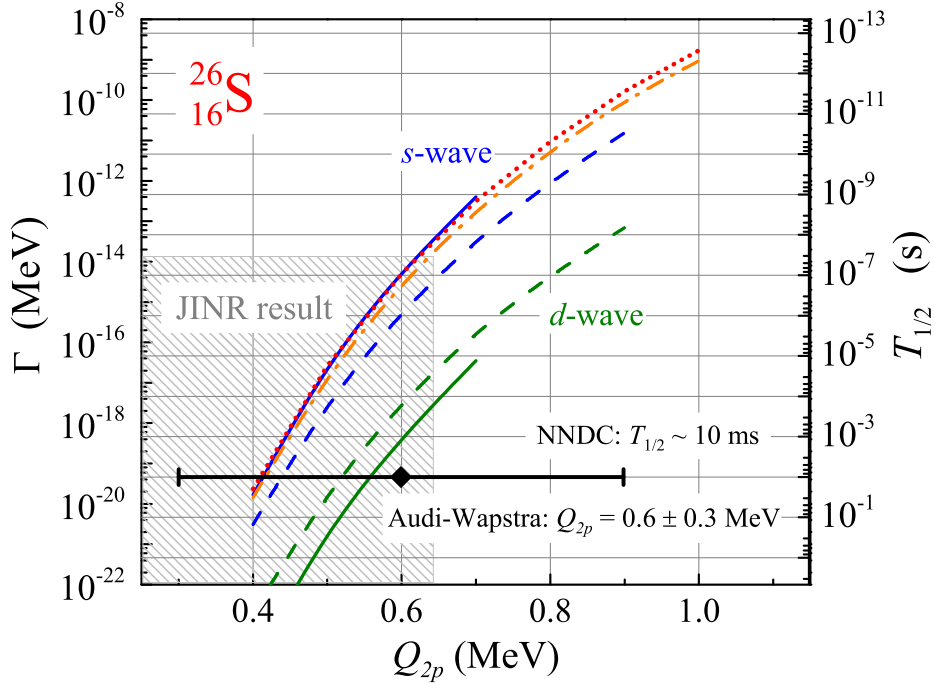}   
}
\caption{(Colour online) Width of the $^{26}$S g.s.\ as a function of the $2p$
decay energy. Solid and dashed curves correspond to the quasiclassical
simultaneous emission model and to the three-body ``$l^2$'' model, respectively.
The RMF-assisted three-body model, based on the relativistic Hartree (RH) and on
the complete RH+BCS results are shown by the dotted and by the dash-dotted
curves, respectively. The lower hatched area shows the energy and the half-life
ranges excluded by the current experiment.}
\label{fig:lifetime}
\end{figure}

For the ground states of both parent and daughter nuclei we apply the
Bardeen-Cooper-Schrieffer (BCS) approximation. In this model the ground state
WF of an even-even spherical nucleus reads:
\begin{equation}
|\Psi_{K}\rangle = |K; \text{BCS} \rangle = \prod_{k >\ 0} \bigl(
u_{(k)} + v_{(k)} a^{\dagger}_k a^{\dagger}_{\bar k}\bigr) |-
\rangle \,,
\end{equation}
where $|-\rangle$ is the bare vacuum. In the case of time-reversal symmetry
for each state $k > 0$, there exists a corresponding time-reversed state ${\bar
k} < 0$ so that all the states $\{k,{\bar k} \}$ together form the whole
single-particle space. The spectroscopic amplitude defined in Eq.\ (\ref{eq:sa})
takes the form:
\begin{equation}
\tilde{X}_{(k)} = B_{(K_i K_f,k k)}^{0} = \sqrt{\Bigl(
j_{k} + \frac{1}{2} \Bigr )} \, u_{(k)}^{(Z-2)}v_{(k)}^{(Z)}\,.
\end{equation}


\section{Structure of $^{26}$S in RMF}


In the present work we employ a RMF model based on the covariant NL3 density 
functional \cite{lal97}. Numerous calculations with NL3 parametrization have 
demonstrated excellent agreement with data in nuclear masses, deformation 
properties \cite{lal97} and properties of giant resonances and nuclear low-lying 
response \cite{lit08}. In RMF model the single-nucleon wave function is a
Dirac spinor characterized by the set $k=\{(k),m_{k}\}$ and
$(k)=\{n_{k},j_{k},\pi_{k},{\tau}_{k}\}$ with the radial quantum number $n_{k}$,
angular momentum quantum numbers $j_{k},m_{k}$, parity $\pi_{k}$ and isospin
${\tau}_{k}$, see Ref. \cite{gam90} for details. For the pairing interaction we
use a monopole-monopole ansatz. The BCS gap equation is solved self-consistently
with the RMF set of equations as described in Ref.\ \cite{lit08}. The strength
of the pairing interaction was adjusted in such a way that the binding energy of
$^{24}$Si and the phenomenological proton pairing gap in $^{26}$S are well
reproduced.

The results of the RMF calculations important for $2p$ decay are summarized in
Table \ref{tab:s26-info} and Fig.\ \ref{fig:uv-evol}. The figure demonstrates
evolution of the $^{26}$S structure with the scaling of the pairing gap. In the
relativistic Hartree (RH) case ($\alpha = 0$) only the nearest to the Fermi
surface $s_{1/2}$ single-particle configuration contributes to $2p$ decay and
the renormalization of the width is trivial ($N_{2p} = 1$). For realistic values
of the pairing gap  ($\alpha \sim 1$) the valence configuration structure for 
$^{26}$S is characterized by the strong configuration mixing with a dominance of 
the $[d_j^2]_0$ configurations.

\begin{figure}
\centerline{
\includegraphics[width=0.9 \textwidth]{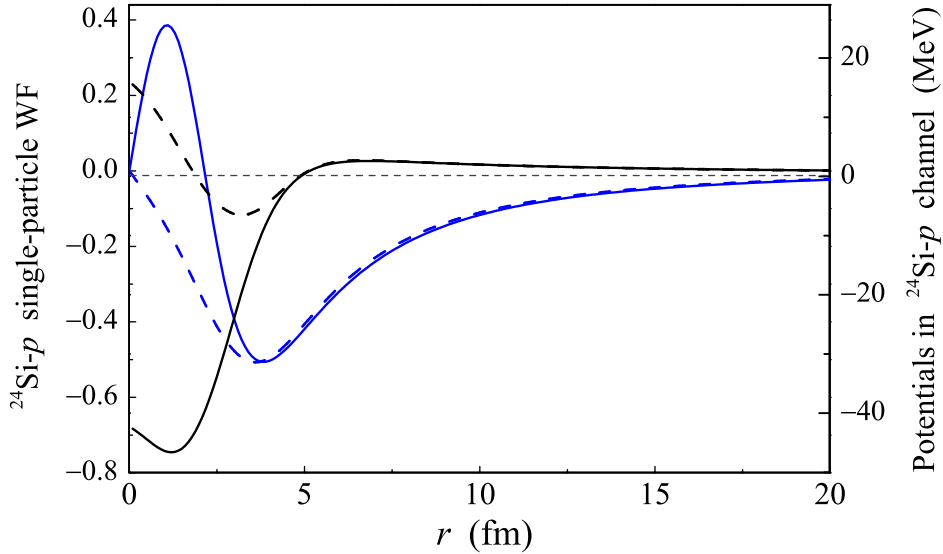}  
}

\caption{(Colour online) Black solid line shows the scaled mean field potential
reproducing the proton separation energy (right axis). Black dashed line shows
the potential with a repulsive core excluding the $s$-wave forbidden state.
Single-particle $s$-wave WFs (found on resonance) are shown for these potentials
by the blue curves of the corresponding style (left axis).}

\label{fig:sp-wf}
\end{figure}


\section{Three-body $2p$ decay calculations}
%

The simple half-life estimates for $^{26}$S are provided in Fig.\
\ref{fig:lifetime} in terms of the models based on systematic values of
parameters which we have broadly used before. The quasiclassical simultaneous
emission model \cite{gri07} neglecting the interaction between the protons
provides a broad ``belt'' of half-life values between the boundaries
connected to emission from the $[s^2]_0$ (upper width limit) and $[d^2]_0$
(lower width limit) configurations. The three-body ``$l^2$'' model \cite{gri03c}
predicts somewhat narrower belt of possible half-lives (however, it is still as
broad as two orders of magnitude).

The RMF results are utilized to modify the three-body calculations in two ways.
(i) The RMF potentials are used to create the core-$p$ interactions for the
three-body model. (ii) The three-body internal structure of the $^{26}$S is
adjusted to be reasonably close to the structure predicted by RMF with a help of
phenomenological short range potentials in $\rho$ variable acting on the
selected hyperspherical components of the WFs $\Psi_3^{(+)}$.

The RMF potential is fitted by the two Woods-Saxon components with
$V_0=\{26,-90\}$ MeV, $r_0=\{0.8,2.87\}$, $a=\{0.55,0.75\}$ fm ($V_{ls}=-14.3$
MeV, $r_0=2.87$, $a=0.75$ fm), see Fig.\ \ref{fig:sp-wf}. The depth of this
potential is scaled to get the RMF positions of the $p$- and $d$-waves in the
nonrelativistic Schr\"{o}dinger equation. For the $s$-waves the modified
potential is used $V_0=\{32.7,-19\}$ MeV, $r_0=\{1.5,4.15\}$, $a=\{0.8,0.63\}$
fm, which contains the core and thus excludes the Pauli forbidden deep
$s$-state. The $2p$ width is very sensitive to the WF behaviour on the nuclear
surface. It can be seen in Fig.\ \ref{fig:sp-wf} that the surface behaviour of
the WF in the modified potential is practically identical to that of the initial
potential. The Coulomb potential in the core-$p$ channel is derived using the
charge distribution of $^{24}$Si obtained in the RMF calculations. The simple
single Gaussian $s$-wave potential $V(r)=-31 \exp[-(r/1.8)^2]$ is used in the
$p$-$p$ channel.

Thus, the difference of the RMF-assisted three-body results shown in Fig.\ 
\ref{fig:lifetime} compared to results of our ordinary ``$l^2$'' calculations 
has the following three sources. (i) The core-$p$ interactions are not taken 
from systematics [$r_0=1.2(A_c+1)^{1/3}$ fm] but are constructed using the RMF 
results. The width is somewhat increased by this choice of the core-$p$ 
interactions. (ii) The three-body structure of $^{26}$S is adjusted to the 
relative weights of the two-proton overlap functions obtained in the RMF 
calculations. This leads to the width reduction compared to the pure three-body 
calculations. (iii) The $2p$ width value should be renormalized according to the 
value $N_{2p}$ of the two-proton overlap functions obtained in the RMF 
calculations. For the predicted structure of $^{26}$S this leads to the increase 
of the width by a factor of two. The latter two trends partly compensate each 
other. Thus, at least for the case of $^{26}$S decay, taking the many-body 
structure of the $2p$ emitter into account leads to the results which are less 
sensitive to model assumptions. The predictive ability of the model is therefore 
expected to be higher even for quite a simplistic method of the structure 
treatment used in this work.

\begin{figure}
\centerline{
\includegraphics[width=0.9 \textwidth]{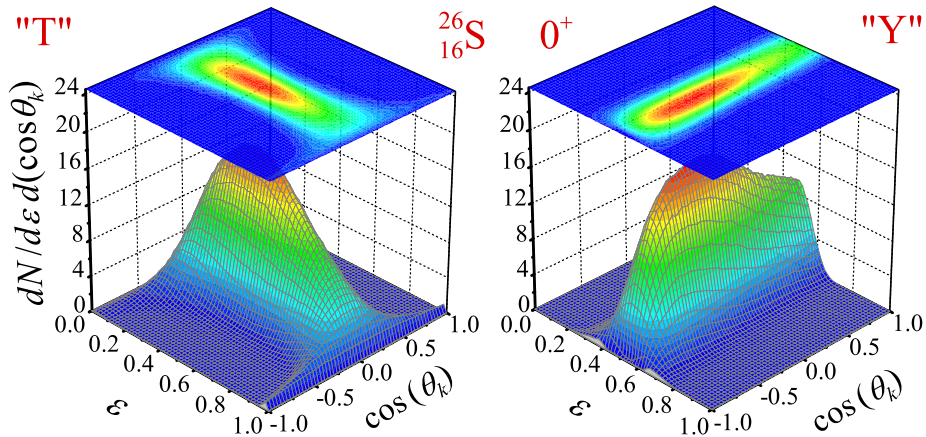}   
}
\caption{(Colour online) Kinematically complete correlation patterns for the 
$2p$ decay of $^{26}$S g.s.\ in the ``T'' Jacobi system ($\varepsilon$ scales 
the relative energy between two protons) and in the ``Y'' Jacobi system  
($\varepsilon$ scales the energy between the daughter nucleus $^{24}$Si and one 
of the protons). Angles $\theta_k$ are defined in Eq.\ (\ref{eq:corel-param}).}
\label{fig:corel}
\end{figure}

This model has important features which we would like to emphasize: (i) the
internal structure of the three-body WF corresponds well to the structure of the
overlaps predicted in RMF (ii) penetration through the Coulomb barrier is
entirely defined by the $p$-$p$ interaction and the single-particle structure of
the core+$p$ subsystems.

Predicted complete momentum correlations for the $2p$ decay of $^{26}$S are
shown in Fig.\ \ref{fig:corel} in terms of the energy distribution parameter
$\varepsilon $ and angle $\theta _{k}$ between the Jacobi momenta 
$\mathbf{k}_{x}$, $\mathbf{k}_{y}$:
\begin{eqnarray}
\varepsilon = E_x/E_T \quad ,\quad \cos(\theta_k)=(\mathbf{k}_{x} \cdot
\mathbf{k}_{y}) /(k_x\,k_y) \, , \nonumber \\
E_T =E_x+E_y=k^2_x/2M_x + k^2_y/2M_y \, , \nonumber \\
{\bf k}_x  =  \frac{A_2 {\bf k}_1-A_1 {\bf k}_2 }{A_1+A_2} \, ,  \,\;
{\bf k}_y  =  \frac{A_3 ({\bf k}_1+{\bf k}_2)-(A_1+A_2) {\bf k}_3}
{A_1+A_2+A_3}\,. \;
\label{eq:corel-param}
\end{eqnarray}
where $M_x$ and $M_y$ are the reduced masses of the $X$ and $Y$ subsystems (see, 
e.g.\ \cite{gri09} for details). The correlation patterns demonstrate the 
regions of strong suppression, connected with the repulsive core-proton Coulomb 
interaction [$\varepsilon \sim 0.5$, $\cos(\theta_k) \sim \pm 1$ in the ``T'' 
system]. Another feature is a broad peak [$\cos(\theta_k) \sim 0$ in the ``T'' 
system] connected with the large weight of the $s$-wave configuration in 
$^{26}$S. Similar correlation patterns have been obtained for other $s/d$-shell 
$2p$ emitters \cite{gri03c} and were later found to be in agreement with the 
experimental data in Ref.\  \cite{muk08}.

Results of the theoretical studies indicate that the considerations for further 
experimental searches for $^{26}$S are especially complicated. If we take the 
standard energy prescription made by Audi and Wapstra \cite{audi} then we have a 
$50\%$ ``probability'' that $^{26}$S can be studied in an implantation 
experiment (half-life exceeding $\sim 100$ ns -- the typical flight-time for the 
fragment separation). Otherwise the decay-in-flight technique \cite{muk07} can 
be used for the $^{26}$S study down to half-life times as short as some 
picoseconds. The $Q_{2p}$ predictions made by different theoretical models point 
at higher energies $Q_{2p} \gtrsim 0.8$ MeV but do not seem to provide a solid 
basis for the choice of experimental technique. We decided to make the first 
attempt to study $^{26}$S in the implantation experiment.


\section{Experiment}
%

\begin{figure}[tb]
\centerline{\includegraphics[width=0.75 \textwidth]{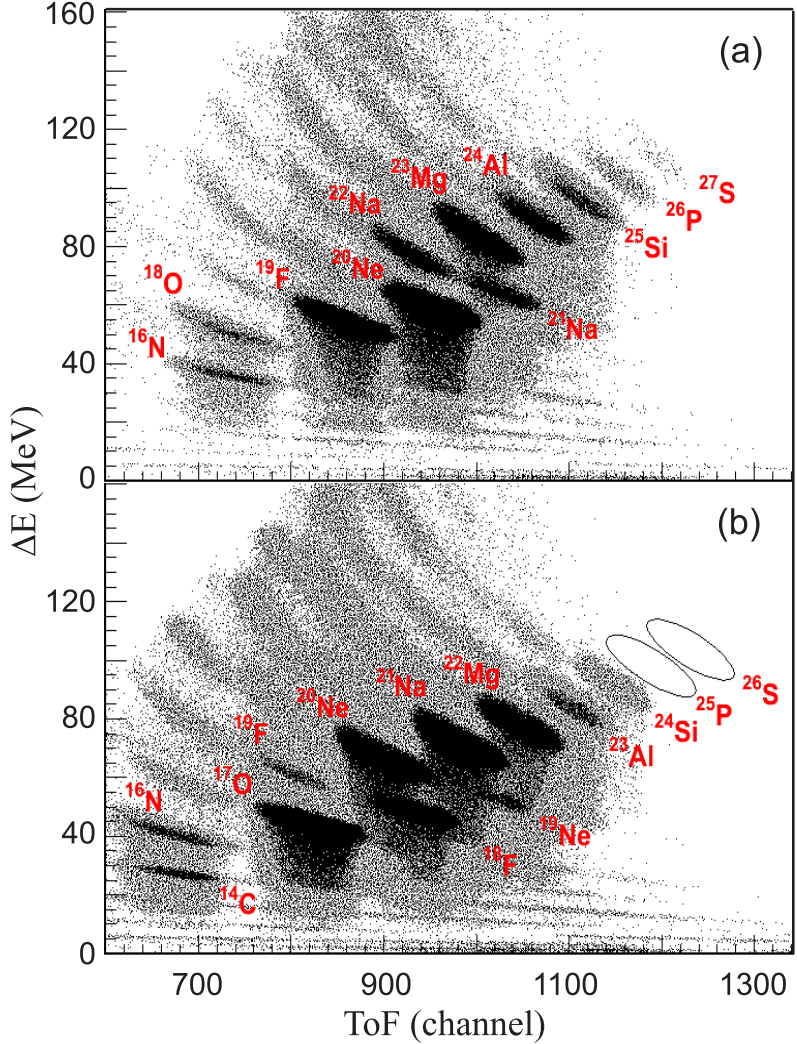}}
\caption{(Colour online) $\Delta E$-ToF identification plots for the cases of
separator tuning on $^{27}$S (a) and $^{26}$S (b).}
\label{fig:expdata}
\end{figure}

The experiment was performed at the ACCULINNA fragment separator \cite{rod97}
(JINR, Dubna, Russia). A 50.3 A MeV beam of $^{32}$S$^{11+}$ ions delivered by
the U400M cyclotron with intensity up to 30 pnA bombarded a 92.4-mg/cm$^{2}$
beryllium production target installed in the separator object plane. A
92.4-mg/cm$^{2}$ beryllium wedge was installed in the intermediate focal plane
of the separator. Combined with a slit restricting the momentum acceptance to
about $2.0 \%$, this rather thick wedge reduced in number the set of $^{32}$S
fragmentation products arriving at the achromatic focal plane when the separator
was tuned for a desired product with its specific charge $Z$ and mass $A$.
Mostly, the nuclei arriving in the beam cocktail at the final focal plane of the
separator were severely depleted in intensity. This provided favourable
conditions to detect $^{26}$S anticipated to be a very rare product of the
$^{32}$S fragmentation. It turned out that only a few fragmentation products
with different $Z$ and $A$ occurred to be close in transmission probability to
the searched nucleus. This appeared to be  useful as such ``satellites'' were
used for tuning the separator when rare fragments, such as $^{26}$S, were
searched for.

To ensure unequivocal identification for the nuclei passing through the 
separator we used both the time-of-flight-$\Delta E$ (ToF-$\Delta E$) and 
$\Delta E$-$E$ techniques. The ToF array, having about 0.6 ns time resolution, 
was placed on a 8.5-meter flight base between the achromatic and final focal 
planes of the separator. The ToF detectors were two 125-$\mu$m thick BC-418 
plastics inclined at $45^\circ$ to the separator axis and each plastic viewed by 
its pair of XP2020 photomultipliers. The telescope composed of one 68-$\mu$m and 
one 1000-$\mu$m, $6 \times 6$ cm$^{2}$ silicon strip detectors ($8 \times 8$ 
strips) was installed in the final focal plane. Besides the $\Delta E$-$E$ 
identification the hit positions of the nuclei were measured by this telescope. 
Also, the signals from the telescope $\Delta E$ detectors were used for the 
ToF-$\Delta E$ identification. In all the runs the load of the strip detectors 
did not exceed 500 pps. Taking into account the granularity of these detectors 
such counting rate makes pileups negligible.

\begin{figure}[tb]
\centerline{
\includegraphics[width=0.72 \textwidth]{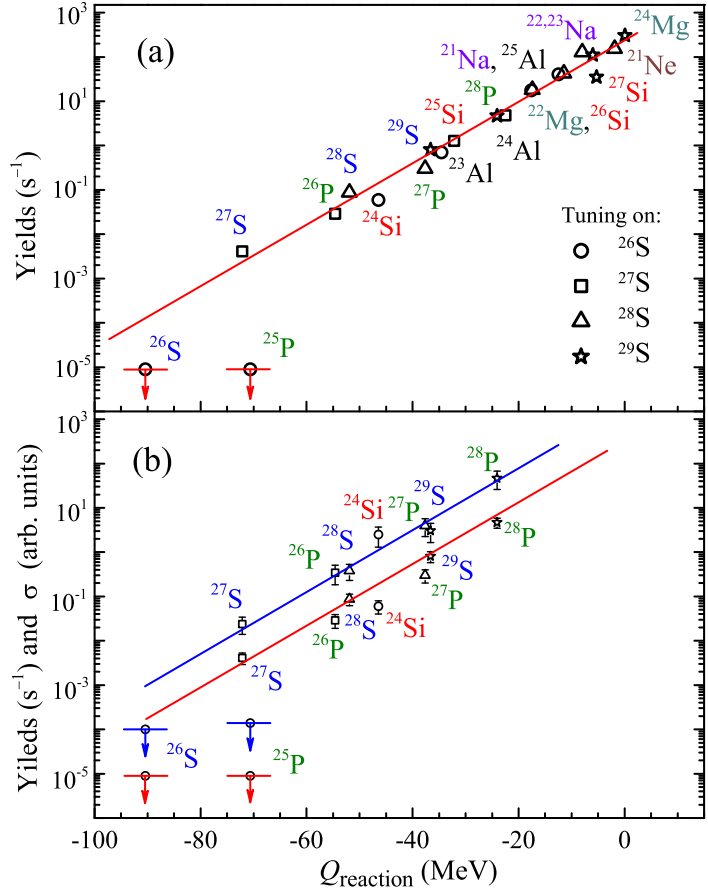}
}
\caption{(Colour online) Panel (a) shows the yields of different nuclei as a
function of the reaction $Q$ value. In this panel the symbol size is
comparable with the error bar values. Panel (b) shows the yields (lower group of
symbols, red line) and production cross sections (upper group of symbols, blue
line) for the isotopes with highest transmission rate through the fragment
separator.}
\label{fig:yield-ot-q}
\end{figure}

Four runs were performed in which the separator was sequentially tuned to give 
the optimum transmission for $^{29}$S, $^{28}$S, $^{27}$S and $^{26}$S. The 
measurements took 870, 1200, 4500 and $1.1 \times 10^5$ seconds, respectively. 
For the two latter cases the $\Delta E$-ToF plots are presented in Fig.\ 
\ref{fig:expdata}. No events of $^{26}$S, as well as $^{25}$P, were found in the 
measurements. To extract the upper lifetime limits for these nuclei, the isotope 
yields of some known nuclei from sulphur to neon were systematized depending on 
their reaction $Q$ values, see Fig.\ \ref{fig:yield-ot-q}. In this figure the 
results of runs tuned for the sulphur isotopes are shown. In addition to the 
three sulphur isotopes ($^{27}$S, $^{28}$S and $^{29}$S) the yields of nuclei 
representing the ``satellites'' in the four separator tunes are presented in 
Fig.\ \ref{fig:yield-ot-q} (a). These satellites are the nuclei having parameter 
$A^{5/2}/Z^{3/2}$ close to that of the tuned sulphur isotopes (variation from 
$-4\%$ to $+3\%$ was allowed). This parameter defines the selectivity of the 
energy loss achromat. For the nuclei selected in such a way the yield dependence 
on the reaction $Q$ values should have exponential character reasonably close to 
the exponential dependence of the production cross sections. Both the yield and 
the cross section dependence on $Q_{\text{reaction}}$ is given in Fig.\ 
\ref{fig:yield-ot-q} (b) for a subset of these nuclei having high transmission 
rates through the fragment separator, namely above $0.25\%$. Naturally, these 
appear to be sulphur, phosphorus, and silicon isotopes. The cross sections were 
defined using the transmission rates calculated with the use of relations from 
Ref.\ \cite{win92}.

The main contribution to the experimental errors stems from the systematic
uncertainties of both the flux of the $^{32}$S primary beam and the transmission
through the separator. It was estimated to be about $25 \%$. The extrapolation
of the obtained systematics shows that in the presented experiment one could 
expect the observation of $16^{+6}_{-4}$ and $360^{+85}_{-70}$ events of 
$^{26}$S and $^{25}$P, respectively. The total
length of the ACCULINNA separator from the production target to the final focus
position is 21.3 m. The full times of flight for these nuclei from the
production target to the detector system  were $314$ ns and $322$ ns (accounting
for energy losses in the wedge and ToF-plastics). Using the Poisson distribution
the half-life limit for $^{26}$S can be found as $T_{1/2}(^{26}$S$)<79$ ns with
$63\%$ confidence level and $T_{1/2}(^{26}$S$)<157$ ns with $98\%$ confidence
level. In these estimations one and four events respectively were assumed
instead of observed zero events. The 79 ns half-life limit is in qualitative
agreement with nonobservation of $^{26}$S as an expected accompanying nucleus in
the recent experiment with a $^{24}$Si secondary beam at MSU \cite{gad08}.

The corresponding half-life limits for $^{25}$P are found as
$T_{1/2}(^{25}$P$)<38$ ns and $T_{1/2}(^{25}$P$)<50$ ns. The $38$ ns half-life
limit is compatible with the limit of $30$ ns provided in Refs.\
\cite{nndc,audi,fir09} (this value seems to be based on an unpublished work). It
can be easily estimated that such a limit provides quite a relaxed limit for the
proton decay energy in $^{25}$P: $Q_p(^{25}$P$) \gtrsim 110$ keV (according to
an R-matrix estimate with systematic parameters). This is not a restrictive
result to define the character of the $^{26}$S decay (true or sequential $2p$
decay, see Table \ref{tab:s26-info} and relevant discussion in the
introduction). However, it helps to provide a two-proton decay energy limit
for $^{26}$S.

Using the dependence shown in  Fig.\ \ref{fig:lifetime} we find that a half-life 
limit of $79$ ns provides a $2p$-decay energy limit $Q_{2p}(^{26}$S$)>640$ keV 
assuming the true two-proton decay. If the one-proton decay were possible for 
this nucleus (which we find to be unlikely situation, but cannot exclude 
completely) one would get a limit for one-proton decay energy 
$Q_p(^{26}$S$)>120$ keV (R-matrix estimate). Combining it with $Q_p(^{25}$P) 
limit we obtain a limit $Q_{2p}(^{26}$S$)>230$ keV. The previously reported 
half-life value of 10 ms \cite{nndc} belongs to the timescale where the decay 
can be attributed to the weak transitions as well (nucleus with such a half-life 
could appear to be even nuclear stable). The half-life times in the nanosecond 
range and shorter would undoubtfully identify the $^{26}$S nucleus as a $2p$ 
emitter.

\section{Conclusions}
%

We have studied both theoretically and experimentally properties of the proton 
dripline nucleus $^{26}$S. Structures of $^{26}$S and its subsystems ($^{24}$Si, 
$^{25}$P)  were obtained in the RMF approach. The $2p$ decay properties of 
$^{26}$S were obtained in a three-body model tuned by the RMF results. It is 
demonstrated that in this joint approach the results appear to be noticeably 
different than the results of the  three-body ``$l^2$'' model based on a 
systematics input. The uncertainties of the $2p$ width connected with structure 
of $2p$ emitters are significantly reduced in this approach.

Production of $^{26}$S in the fragmentation reaction was studied experimentally.
We have found that either the production of $^{26}$S is anomalously small (more
than one order of magnitude than expected according to systematics) or the
lifetime is too short. The systematical value of the production cross section
enables us to estimate the upper half-life limit of $^{26}$S as 79 ns. 
Theoretical lifetime
dependence on the $2p$-decay energy allows one to infer the lower limit of 640
keV for the $Q_{2p}$ of $^{26}$S. This value is in principle consistent with the
theoretically predicted range of mass excess for $^{26}$S but it narrows
significantly its uncertainty. In the case that $^{26}$S has an open one proton
decay branch (sequential two-proton decay) a less restrictive limit $Q_{2p}>230$
keV is obtained.

Our results are in contradiction with the existing information about $^{26}$S as
a possible $2p$ emitter with the half-life of about 10 ms \cite{nndc}. Our
results allow focusing the search for $^{26}$S in the half-life range 10 ps --
100 ns where the decay-in-flight technique \cite{muk07} is the adequate
experimental method.

%
\section*{Acknowledgements}
%

The authors are grateful to Profs.\ Yu.Ts.\ Oganessian and S.N.\ Dmitriev for
the overall support of this experiment and Prof.\ M.\ Pf\"{u}tzner for
illuminating discussions. This work was supported by the Russian Foundation for
Basic Research grant  RFBR 08-02-00089-a. L.V.G.\ is supported by Deutsche
Forschungsgemeinschaft grant 436 RUS 113/907/0-1, FAIR-Russia Research Center
grant, RFBR 08-02-00892, and Russian Ministry of Industry and Science grant
NSh-7235.2010.2. E.V.L.\ acknowledges support by the Hessian LOEWE initiative
through the Helmholtz International Center for FAIR and by the Russian Federal
Education Agency Program. V.C.\ was partly supported by Czech grant LC07050.



\end{document}